# A highly scalable fully non-blocking silicon photonic switch fabric

Dessislava Nikolova[1]*, David M. Calhoun[1]*, Yang Liu[2], Sebastien Rumley[1], Ari Novack[1,2], Tom Baehr-Jones[2], Michael Hochberg[2], Keren Bergman[1]

[1]Department of Electrical Engineering, Columbia University, 530 West 120[th] street, New York, NY 10027, USA

[2]Coriant Advanced Technology Group, 171 Madison Avenue, New York, NY 10016, USA

*Both authors contributed equally to this work

Corresponding author: Dessislava Nikolova dnn2108@columbia.edu, tel: +1-212-854-8068  Fax: +1-212-854-2900

David Mark Calhoun dmc2202@columbia.edu,

Yang Liu yhvhliuyang@gmail.com,

Sebastien Rumley sr3061@columbia.edu,

Ari Novack arinovack@gmail.com,

Tom Baehr-Jones tbaehrjones@gmail.com,

Michael Hochberg michael.hochberg@gmail.com,

Keren Bergman kb2028@columbia.edu,

**ABSTRACT**

Large port count spatial optical switches will facilitate flexible and energy efficient data movement in future data communications systems, especially if they are capable of nanosecond-order reconfiguration times. In this work, we demonstrate an 8x8 microring-based silicon photonic switch with software controlled switching. The proposed switch architecture is modular as it assembles multiple identical components with multiplexing/demultiplexing functionalities. The switch is fully non-blocking, has path independent insertion loss, low crosstalk and is straightforward to control. A scalability analysis shows that this architecture can scale to very large port counts. This work represents the first demonstration of real-time firmware controlled switching with silicon photonics devices integrated at the chip scale.

**INTRODUCTION**

Optical switching has the potential to enable ultra-high capacity optical networks that can deliver large volumes of data with time-of-flight latencies [1,2,3]. Silicon photonic technology—with its CMOS compatibility, compact footprint, high bandwidth density, and the potential for nanosecond scale dynamic connectivity—offers a promising platform to realize optical switches [4,5]. Large port count micro-electro-mechanical switches have been demonstrated; however, such devices do not achieve nanosecond scale dynamicity, require high actuation voltages that are not suitable for all applications [6]. Silicon photonic switching elements employing Mach-Zehnder interferometers or microring resonators provide 1x2 and 2x2 spatial switching enabled by thermo-optic or electro-optic actuation, with typical applied voltages close to 1V. By combining multiple 1x2 or 2x2 switching elements port scaling can be realized. This scaling of nanophotonic switches on a 2D plane is primarily limited by the losses and crosstalk induced by active switching elements and passive interfaces such as waveguide crossings [7,8,9,10].

A desirable property for a switch fabric is path-independent insertion loss (PILOSS) [11]. A switch that exhibits PILOSS has identical insertion loss on all paths through the switch. The blocking characteristic of the switch is also a crucial determinant of the switch performance. Blocking switches do not guarantee simultaneous transmission from all inputs even if they all have different output port destinations. Reconfigurably non-blocking switches might have to interrupt ongoing transmissions to satisfy an alternative connection requirement [12], while blocking switches simply cannot satisfy some requirements. In a fully (or strictly) non-blocking architecture, the paths from any input port to any output port can be used simultaneously, resulting from the state of the switching elements on a given path being independent from the state of all other switching elements. Fully non-blocking switch fabrics without PILOSS can be realized based on a crossbar topology [13] and with a switch-and-select architecture [14]. The switch and select architecture consists of N number of 1-to-N input switch arrays, N number of N-to-1 output switch arrays, and a total of $N^2$ on-chip waveguide crossings. In this architecture, the large number of on-chip waveguides limits the potential for scalability and introduces different insertion loss on the different paths. Fully non-blocking switch fabrics on a single chip with PILOSS have been realized [15-17]. The number of switching elements required for these architectures scales with the square of the number of ports.

System-level implementation of silicon photonic devices hinges on seamless integration with traditional computing hardware and software (i.e. operating system, protocols, etc). Custom-designed firmware with the ability to control the switch at a fine level of granularity—subsequently, offering more advanced network functionalities—is required [18]. Hardware-software integrated silicon photonic subsystems controlled by means of custom firmware implemented in field programmable gate arrays (FPGA) have been demonstrated implementing simultaneous control of two rings on one chip [19, 20].

In this work we propose a silicon photonic switch fabric with PILOSS and fully non-blocking configuration capability. We demonstrated a proof-of-concept for 8x8 switch fabric by assembling multiple silicon photonic integrated circuits (PIC) offering multiplexing and demultiplexing functionalities. Our analyses show that the proposed architecture can be scaled up to high port-counts even with current state-of-the art silicon photonic devices. It requires only two switching elements per port active at a time, which can potentially lead to very low power consumption. We have implemented software-controlled switching functionality, demonstrating the feasibility of a readily available control plane that can be scaled to even more advanced networking utilities. This easy-to-realize, easy-to-control and modular architecture can lead to the faster adoption of silicon photonics for optical switches in data communications.

**METHODS AND MATERIALS**

**Switch architecture**

The proposed switch (Figure 1) consists of multiple silicon photonic integrated circuits (PIC) interconnected via off-chip cross-connects. Each separate switch input interface is a PIC with a spatial demultiplexer (demux), and each output interface is realized as a PIC with a spatial multiplexer (mux). At each input interface, the optical signal is coupled to the demux PIC acting as a spatial 1-to-N switch. The demultiplexing functionality can be realized with N microrings, coupled to a bus waveguide as shown on Figure 1.The microrings act as the basic switching element directing an input signal to either the add or drop port (Figure 1, right-top) [21] One side of the bus waveguide is the ingoing port and the microrings' drop ports are the chip's outgoing ports. These microrings can be tuned with an applied voltage to couple light off of the bus waveguide to a drop port. Each microring's drop port is an outgoing port of the demux chip, and is connected through an off-chip fiber to the ingoing ports of the output interface PIC. Signals are coupled onto the mux PIC via add port waveguides and microrings, and subsequently onto the bus waveguide of the mux and to the mux's outgoing port. By coupling the signal from the input interface and guiding it to the output interface through optical fibers, the need for on-chip waveguide crossings is mitigated at the cost of chip-to-chip insertion loss. By introducing active switching elements on the output interface via a mux, this architecture aims to significantly reduce inherent architectural constraints of a cross-connect such as crosstalk and subsequent switch scalability.

The ubiquity of such a microring-based architecture is underlined in the fact that the mux and demux PICs are one and the same, simply used by propagating light in opposite directions.

In order to ensure that the insertion loss and crosstalk bounds are the same for each input-output pair, the connectivity between the PICs is made such that the signal from any input port to any output port passes through exactly N+1 microrings in total. In other terms, the drop port at index i must always connect to the add port at index (N+1-i). This PILOSS connectivity is shown in Figure 1: input demux 1 on outgoing port 1 is connected to ingoing port 8 on output mux 8, and the signal traverses 9 total microrings chip-to-chip.

**Architectural evaluation**

The microrings used for spatial switching in the implemented architecture are especially suited for this type of switch where the switch elements do not require the same loss to switch the signal from one path to another. Consider a microring between two waveguides (bottom-right inset in Figure 1): in "bar" state, the input optical power $P_{in}$ propagating along a waveguide will be transferred further along the same waveguide with some loss due to the coupled microring $P_{through} = T\, P_{in}$. In "cross" state, part of the power $P_{in}$ will couple to the microring and then to the drop waveguide, $P_{drop} = D\, P_{in}$. Each drop port provides a filtered version of the input signal according to the shape of the resonance, which for a single microring is a Lorentzian.

The through coefficient T when the microring is exactly tuned between two resonances and drop coefficient D when the ring is on resonance are given with $T = \frac{(1+a)^2 t^2}{(1+t^2 a)^2}$ and $D = \frac{(1-t^2)^2 a}{(1-t^2 a)^2}$, where $a = e^{-\alpha L/2}$ is the optical field loss of a microring with circumference L and optical power loss coefficient α. The coupling coefficient, *t*, indicates the portion of the incoming electrical field on a waveguide that continues to propagate further along after a microring [22]. *t* depends on the geometry of the coupling region [23].

The optical signal in the proposed architecture has to be coupled into and out of two chips before being detected on a receiver; subsequently, it must pass through a total of N+1 microrings: N-1 microrings in bar state and 2 microrings in cross state. Furthermore, it has to be guided in an on-chip waveguide, which exhibits a distance-dependent loss $W_{IL}$. Taking into account these losses on the optical path the signal at the output port is given with:

$$P_{signal} = P_{in}C_{IL}^4 W_{IL} D^2 T^{N-1}, \qquad (1)$$

where $C_{IL}$ is the loss per coupler,. The leaked power to any output port $P_X$, is:

$$P_X = P_{in} C_{IL}^4 W_{IL} ((K-1)(1-T)^2 T^{N-1} + (N-K)(1-T)^2(1-D)T^{N-2}) \qquad (2)$$

where K can range from 1 to N-1 depending on the destination port of the particular input contributing to the crosstalk For the case (1-D)<= T the maximum total crosstalk power penalty [24] becomes:

$$PP_{Xtalk} = -10Log(1 - \sum_i 2\sqrt{P_{X,i}/P_{signal}}) \qquad (3a)$$

$$\leq -10Log(1 - 2(N-1)(1-T)/D). \qquad (3b)$$

**Device characterization**

To demonstrate the feasibility of the proposed architecture, we use two 8-channel PICs. A signal being switched through the proposed switch architecture traverses two PICs—at the input and output (I/O) interfaces—and is completely unaffected by the state of the microrings on PICs of other I/O interfaces. The specific PICs used in this work are fabricated at the Institute of Microelectronics (IME)/A*STAR, Singapore via an OpSIS multiple-project-wafer run. The process starts with 8 inch Silicon-on-insulator (SOI) wafer with 220 nm top silicon layer. Three dry etch steps are used to define the silicon microrings and grating couplers. Six implantation steps are applied to form the heater and contact region in the silicon microrings. Two levels of aluminum are deposited for electrical interconnection. Each 8-channel mux/demux consists of eight silicon microring resonators, side-coupled to a bus waveguide and each coupled to an individual waveguide. The transmission characteristic at the through ports of the bus waveguides of the two chips is shown in Figure 2b. Without any tuning, the ambient resonances of the eight microrings have different frequencies, so the operating wavelength should be such that all ambient microring resonances do not interfere. The free spectral range (FSR) of a microring is approximately 13.2 nm. The microring is thermally tuned by an integrated heater. The required applied voltage to shift the resonance a full FSR is approximately 4.2V. The change of the resonance with the applied voltage for microring 7 on chip 1 (demux) is shown on Figure 2c. Figure 2d shows the same microring on chip 2 (mux). The cascaded effect of the other microrings along the waveguide causes the observed resonance shape to be asymmetric. The microring on chip 1 is heated, while the microring on chip 2 is cooled, causing red-shifting and blue-shifting in resonance, respectively. Cooling was achieved by reducing the applied the voltage on the ring relative to its initial value, which was applied to shift its resonance with one FSR. Heating the microring to move it away from resonance results in higher power drop for the same applied voltage than when the microring is cooled due to the asymmetry of the resonance.

## RESULTS

**Switching devices and architecture**

Three separate experimental approaches were used to demonstrate the efficacy of the chip-to-chip microring cross-connect as a switching architecture with system-level characteristics. These approaches primarily focused on architectural characterization, crosstalk characteristics, and a system-level switching demonstration.

The precise connectivity between the ports required to realize an 8x8 switch—ingoing port 1 of demux 1 is connected to ingoing port 8 of mux 1; port 2 of demux 2 is connected to port 7 of mux 1; etc.—is emulated according to Figure 3a. Measuring the insertion loss through the proposed switch is therefore sufficiently demonstrated using only two PICs. Figure 3b shows the output power for the different possible input-output paths between two chips obtained by connecting different ingoing to outgoing ports. For each measurement only one of the connections shown on Figure 3a is realized. The upper row markers, show the insertion loss measurements from the first chip on each of its outgoing ports. The polarization of the input signal was set by maximizing the output power from the bus waveguide while passing by all microrings tuned in the bar state. There is a slight decrease in the output power towards higher number of the demux outgoing ports as the insertion loss increases when the signal is drop from ports situated more towards the end further from the ingoing port. The middle row markers are the measured insertion loss on the output of the second chip for each ingoing-outgoing port combination as shown on Figure 2a after the signal has passed through 7 microrings tuned away from resonance (bar) and 2 microrings tuned on resonance (cross). Variations in insertion loss on different paths can be attributed to several factors: manufacturing tolerances directly affecting microring size, the geometry of the coupling region between straight waveguides and microrings, and the I/O grating couplers. By tuning the polarization or the applied voltage on the microring these variations can be minimized and the PILOSS enforced. In our experiment the polarization of the signal between the paths was tuned such that the output power amongst the different paths was completely equalized, inducing a scenario where each path experiences the worst case insertion loss.

The bottom row markers show the measured output power for the different outgoing-ingoing port combinations when the connected microrings are tuned close to half a FSR away from resonance. The power dropped by the microrings while still in bar state is the crosstalk power each path will contribute to a signal. The polarization of the signal is the same as the polarization that achieves the desired insertion loss. The realized architecture provides excellent port isolation—close to 39 dB.

This low crosstalk power is expected to result in a negligible power penalty to the signal. To demonstrate the crosstalk power penalty, we have measured the bit-error rates (BER) of a signal when a varied number of signals contribute to the added crosstalk. To realize this, we utilized the experimental setup as shown on Figure 4. The signal before being coupled to the first chip is split in two: one portion is used to create the crosstalk and the second portion is the measured signal. A passive 90:10 power splitter was used to tap off a copy of the signal generated by modulating a $2^{31}-1$ pseudorandom sequence on a CW laser at 1549.6 nm. This signal was then attenuated and sent to the mux chip where it was dropped by microring 8. The attenuator on its path is used to control its optical power such that it is the same as the power of a signal passing the demux microring 1 to mux microring 8 path. All other microrings except microring 8 on the second chip were tuned away from resonance using an FPGA-enabled tunable voltage supply.

The 90% of the optical power was coupled onto the demux chip to create the crosstalk signals. The different paths emulate 6 different possible crosstalk paths, consisting of microrings 2 through 7 on the demux connected to 7 through 2 on the mux. All paths are decorrelated by using different fiber lengths in the cross connect. Figure 5b gives the measured and expected power penalty versus the number of crosstalk sources. In order to calculate an expected value for the power penalty we obtained the coupling coefficient t and optical field loss for each ring from the measured through and drop powers [25]. To obtain T for each chip, values for $t$ and $a$ were taken as the average of these values for rings 2 to 7. The values for D for each chip is obtained from the estimated $t$ and $a$ for ring 8 for the first chip and ring 1 for the second chip. The theoretical bounds on the power penalty were subsequently obtained from Eq. (3b). The measured power penalties were extracted from the BER curves shown on Figure 5a as the difference between two curves for nominal error-free BER of $10^{-12}$. The crosstalk powers as indicated on Figure 3b are close to the noise level, and were subsequently observed to fluctuate throughout the experiment. Nevertheless, the theoretical model provides a bound on the measured crosstalk power penalty, which on average remains below the calculated bound. A detuning of the microring resonance will result in less signal power, which will have an effect on the total power penalty. To demonstrate the extent of this effect, the optical attenuator on the signal path was used to measure the BER for different signal powers corresponding to different microring detuning. Figure 5c shows an increase in power penalty according to decreasing signal power, which corresponds to detuning the resonance of one of the two drop microrings on the path from the signal wavelength. The correspondence between detuning and optical power decrease depends on the sharpness of the resonance, which depends on its Q-factor. Microrings with higher Q-factor will be more sensitive to detuning, and methods to control the stability of resonance tuning have been explored [26,27].

**System-level switching demonstration**

We performed an active switching scenario by actuating two microrings on both the input demux and output mux PICs as indicated in Figure 4. A controller circuit board employing an Altera Stratix V field programmable gate array (FPGA) and digital to analog converters (DAC) clocked at 65 MHz was used in conjunction with conventional op-amp electrical level shifters to provide the appropriate voltages to shift each microring's resonance across its full FSR. This direction-actuation schema to tune microrings to particular resonances consisted of custom firmware—a combination of register transfer level (RTL) hardware description language (HDL) to configure the DACs and to form an embedded processor—on top of which embedded software code written in C could be executed. This full hardware-software integrated system was used to demonstrate the system-level switching characteristics of the proposed architecture.

The first switching experiment we performed was meant to characterize the rise and fall times of microrings in this switching architecture. We performed an active switching scenario by concurrently actuating two microrings on both the input demux and output mux PICs as indicated in the inset of Figure 6. Figure 6 shows representative results of the entire system: a pair of microring resonances are tuned by fixed amounts across a demux/mux pair. These resonance shifts correspond to a particular voltage sweep, to and from the voltage required to reach resonance tuned to the signal wavelength. A $2^{31}$-1 pseudorandom binary sequence was modulated onto a continuous-wave laser at 1549.6 nm, and was then amplified and injected into the ingoing port of the demux PIC. The signal from the outgoing port of the mux was coupled off-chip, amplified, and observed on a digital communications analyzer.

While previous works using electro-optic actuation of microring switches show nanosecond and sub-nanosecond rise times [28, 5], the switching setup time due to thermal actuation in this work is on the order of microseconds. This work shows the relationship of tuning the resonance over varying intervals—or distances shifted between on-resonance and off-resonance—with thermal setup times at single-digit microseconds without a closed-loop control system. Both red and blue shifts on the demux and mux were used to average the effects of heating and cooling microrings, with the intention of equalizing rise and fall transients. However, the gradual elongation of rise time as observed in Figure 6a-g can be attributed to the fact that when tuning two microrings on resonance, both need to reach resonance before a completed optical path can be constructed. In the case of the fall time, the optical path is deconstructed as soon as either resonance shifts an appreciable amount. We suspect the effect of overall larger resonance shifts on the demux side primarily caused the elongated rise time; the larger resonant shifts for the demux compared to the mux are attributed to the electrical amplification components used level-shift DAC outputs to appropriate voltage swings in this experiment.

The measured rise time for tuning both PICs to establish an optical path in Figure 6a—the small resonance-shift case—is approximately 1 μs. Figure 6g—the large resonance-shift case—shows an optical path achieved in approximately 10 μs. The resonance-shift in Figure 6 in between the two extreme cases scale accordingly. The order of magnitude difference between the extremes can be attributed to the rate at which the integrated heaters in the silicon substrate disperse current-induced heat to the microring structure. In the small resonance-shift case, the heat differential between on-resonance and off-resonance microring tuning is smaller than for the large resonance-shift case. In both cases, the rise time of the applied electrical signal is several orders of magnitude faster than the resulting optical rise time, which confirms the limitation on thermal actuation properties primarily influenced by the thermo-optic coefficient of silicon [27].

We additionally characterized the switching time when transitioning between the two signal paths, as shown in Figure 7. To realize this experiment we controlled simultaneously 4 microrings over two chips. We observed similar rise/fall transients, but with different total signal power levels for each channel, likely due to a path differential caused by passive off-chip interconnection components (polarization controllers, fibers, couplers, etc.). This demonstration shows successful multipath switching through the switch architecture, with overall switching times that are suitable for microsecond-scale granularity. Additionally, the firmware and software used for switching is easily scalable beyond 4 microrings due to the modular nature of RTL coding and embedded software coding.

## DISCUSSION

**Switch fabric scalability**

The practical insertion loss (~38dB) of the demonstrated switch fabric is significant due to the employed optical grating coupler technology, inducing ingoing and outgoing insertion losses and mode mismatching [298]. However that can be significantly improved with state of the art coupler designs and optimized ring coupling geometry, but is outside the scope of this work. Coupling losses of 1.5 dB per coupler have been reported [16] and low loss waveguides, crossings, and bends are also achievable. An upper bound of the total power penalty from the device is given by $PP_{total}=IL_{dB}+PP_{Xtalk}$, with the insertion loss $IL_{dB}=-10\log(P_{signal}/P_{in})$ [24] where $P_{signal}$ is given with Eq (1) and the crosstalk power is given with Eq. (2). The dependence on the total power penalty of a device on the coupling coefficient t is shown on Figure 8a. The graph is obtained with CL=1.5dB, $WG_{loss}$=1dB/cm and a=0.99; there is a particular coupling coefficient which minimizes the total power penalty. We observed power penalty that remains close to the minimum value over a sufficiently wide range of the coupling coefficient, inferring that the proposed architecture is not highly susceptible to fabrication uncertainties. For very large radices, this range becomes smaller and the power penalty will be more sensitive to deviations from this optimal value. The coupling depends largely on the distance between the waveguide and the resonator which can be set at device design time [29].

Figure 8b shows the expected increase of the total power penalty according to increasing switch radix, calculated using the optimal coupling for each radix. The part of the power penalty which depends on the switch radix hence the number of microrings on the optical path and the coupling t is also shown. The most significant contribution to the total power penalty for radices less than 200 is from the couplers and waveguide losses, which continue to be a focus for improvement for the foreseeable future.

We also indicated power penalty for a switch that does not use a second microring-based output interface. Specifically, we have considered the case where the multiplexer interfaces were not implemented as switching elements but as lossless Y junctions, whose insertion loss and crosstalk characteristics at each output are different. The crosstalk power penalty is given as $PP_{Xtalk,Y} \leq -10Log(1 - 2(N-1)\sqrt{(1-T)/D})$. The power penalty for such switch is significantly higher than for the proposed architecture, showing the extent to which the multiplexer PICs at the switch output ports contribute to signal isolation and the corresponding crosstalk reduction. This result highlights the advantage of the proposed switch design in terms of scalability. It provides high port isolation and as a result the expected power penalty is low and increases slowly with the switch radix. By avoiding on-chip crossings and optimizing the mode coupling this switch architecture reduces crosstalk due to cross-connects, which makes it highly scalable. Using only two active switching elements per input-output port interconnection further lowers the insertion loss as microrings typically have higher loss in the drop state.

**Impact of active switching**

Designing the microrings such that they are off-resonance with the desired wavelengths allows per-port tuning of only two microrings on resonance to achieve port-to-port connectivity. This results in the proposed architecture needing only 2N microrings to be actively tuned for a switching event, hence making it superior in terms of power consumption in comparison with other architectures. Additionally, the use of independent switching subsystems at each input and output makes it possible to realize schemes where if the mux/demux at one port is defective, only this port can be replaced or avoided entirely.

While the time-granularity of switching demonstrated here falls short of the nanosecond scale often cited as a key requirement for various levels of computing, this switching architecture serves as a basis for later improvements that can still take advantage of the architecture's low-crosstalk characteristic. Electro-optic switching offers a straightforward timescale improvement with sub- nanosecond switching time already demonstrated. Other options include co-integration of this thermally actuated architecture with additional photonics, including other interferometric structures such as Mach-Zehnders or fast-tunable laser sources.

A hardware-software integrated control solution offers abstraction of the architecture that is compatible with high performance computing systems. Additionally, modular design of both RTL and embedded software allows for both ease-of-control and easy scaling to a multitude of simultaneously-controlled microring devices. The time-granularity constraint of this demonstration, when combined with a real computing framework, can be amortized with intelligent protocols and system design for maximizing the interconnect utilization on an as-needed basis. In other words, it is conceivable to design a computing framework whose interconnection need is met by the silicon photonic switching architecture shown here, and architectural improvements of this silicon photonic switch offer to further improve future high performance computing interconnects.

## Acknowledgements

The work of David M. Calhoun is supported in part by the Columbia University Optics and Quantum Electronics IGERT under NSF IGERT (DGE-1069240). The authors would like to thank Gernot Pomrenke, of AFOSR, for his support of the OpSIS effort, through the PECASE award (FA9550-13-1-0027), subcontract 39344 Multi-Terabit-Capable Silicon Photonic Interconnected End-to-End System under DURIP (FA9550-14-1-0198), and ongoing funding for OpSIS (FA9550-10-1-0439). This work was further supported in part by the AFOSR Small Business Technology Transfer under Grant FA9550-12-C-0079 and by Portage Bay Photonics.

## Conflict of interest

All authors declare no conflict of interest.

## Author contributions

These authors contributed equally to this work

D. Nikolova and D.M.Calhoun

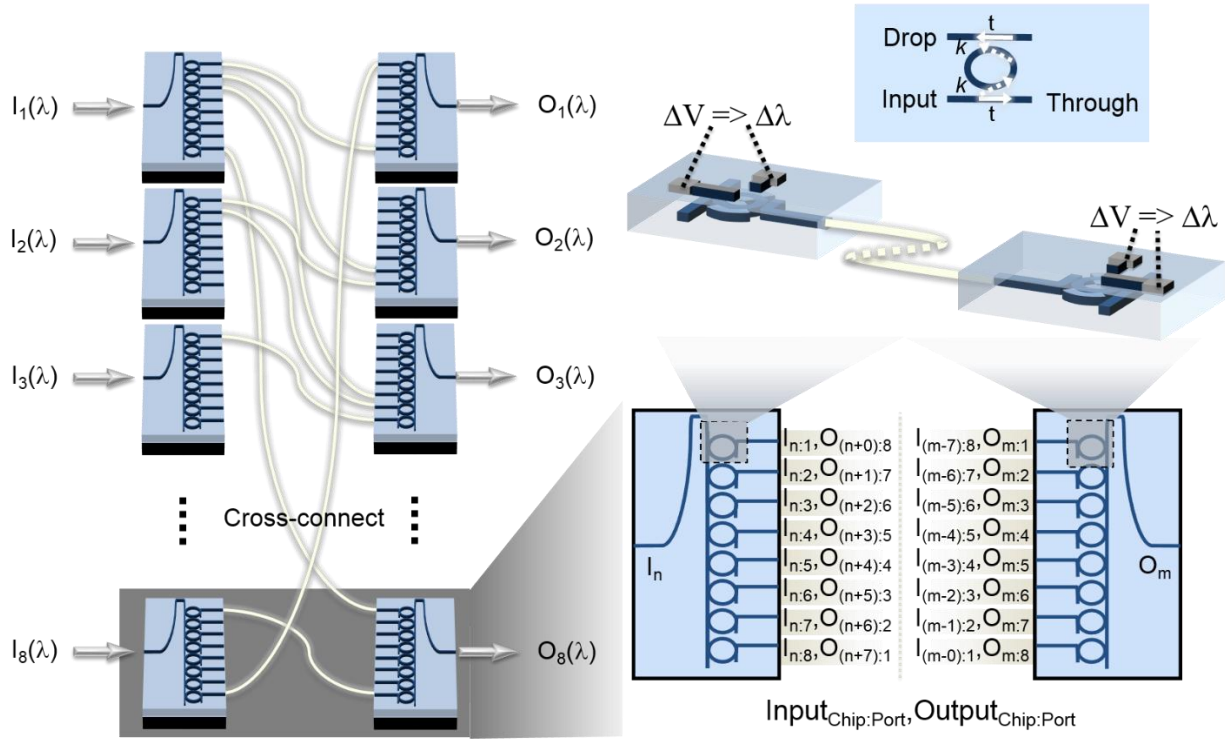

**Figure 1** Schematic representation of input-output connectivity of silicon photonic microring switching architecture; **(left)** depiction of chip-scale integration of 8-microrings multiplexers and demultiplexers to achieve connections between inputs, $I_n$, and outputs, $O_m$; **(right-bottom)** indexed representation of all potential intermediate cross-connects between multiplexers and demultiplexers; **(right-middle)** implanted on-chip heaters consisting of highly-doped regions connected to metallic conductors, inducing a shift in microring resonance proportional to applied power; **(right-top)** a microring between two waveguides acting as a switch of an input signal between the through and the drop ports.

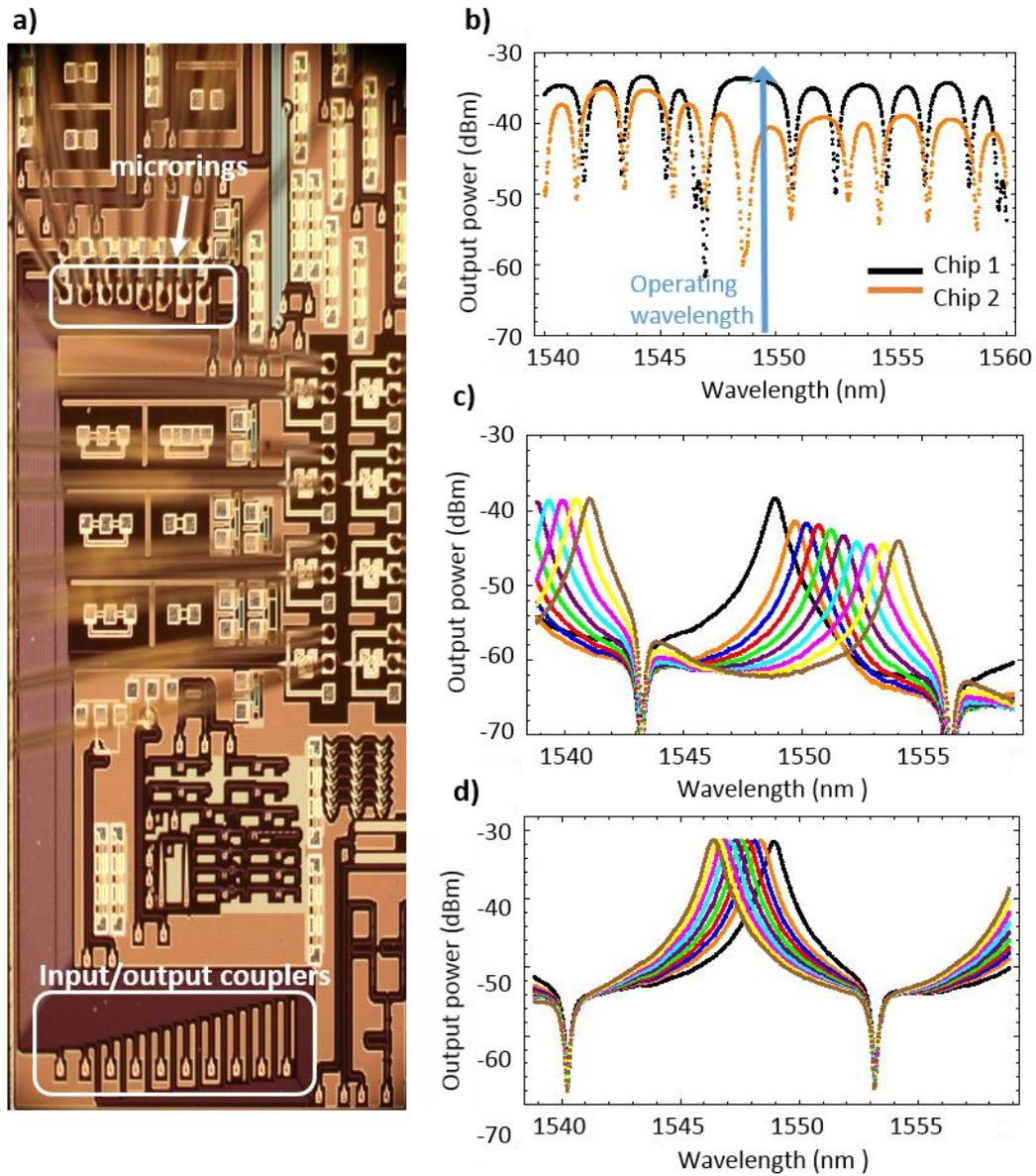

**Figure 2** **(a)** An optical micrograph of one wirebonded PIC used in this work; **(b)** the transmission characteristic between the ingoing to outgoing port of the bus waveguides on the two chips. At the operating wavelength all microrings are off resonance. Tuning the voltage on the thermal heater of a microring (ring 2 in the example) shifts the resonance wavelength and changes the transmitted power to the drop port; **(c)** heating the microring shifts the resonance towards longer wavelength and correspondingly **(d)** cooling the microring (decreasing the applied voltage relative to some set initial value) shifts it to shorter wavelengths.

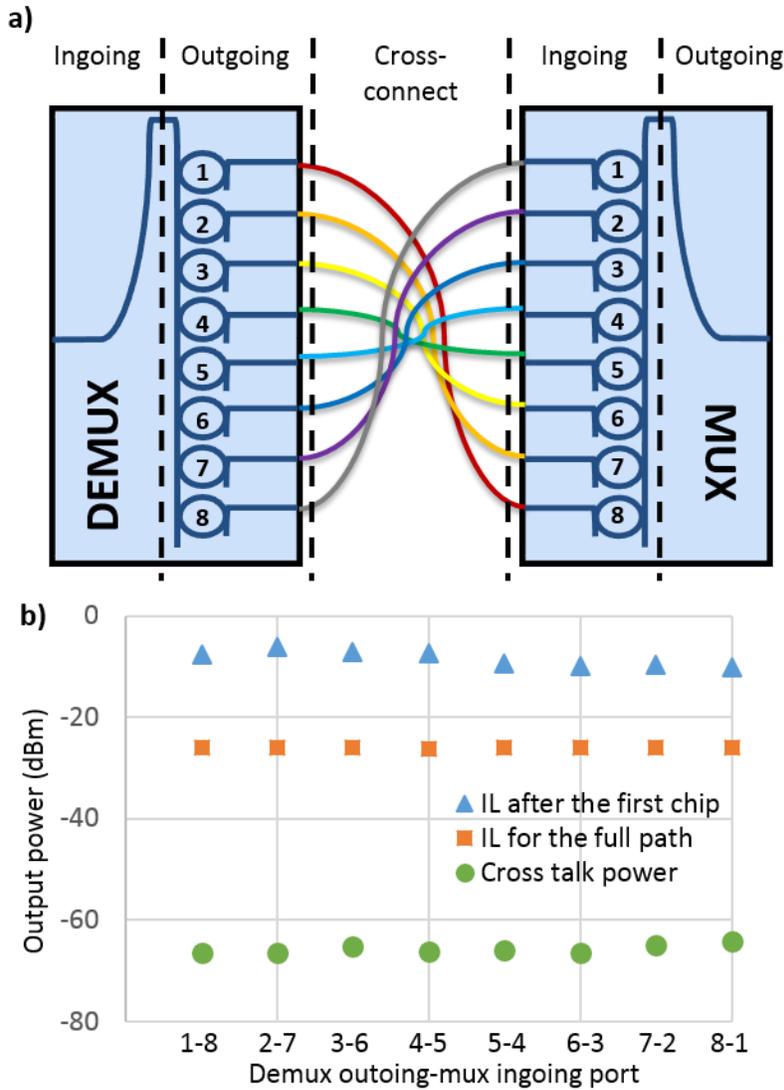

**Figure 3** Architectural evaluation scenario utilizing two PICs to emulate the PIC-related insertion loss between the different in- and out-going ports combinations. **(a)** The ingoing-outgoing port configurations with the different colors representing different input-output ports connectivity; **(b)** the measured power of the signal at the outgoing ports of the first chip, after passing the full path through the switch i.e. at the outgoing pot of the second chip and the corresponding crosstalk power when the rings are tuned a half FSR away from resonance. Only one fiber in the cross-connect was coupled for each separate measurement.

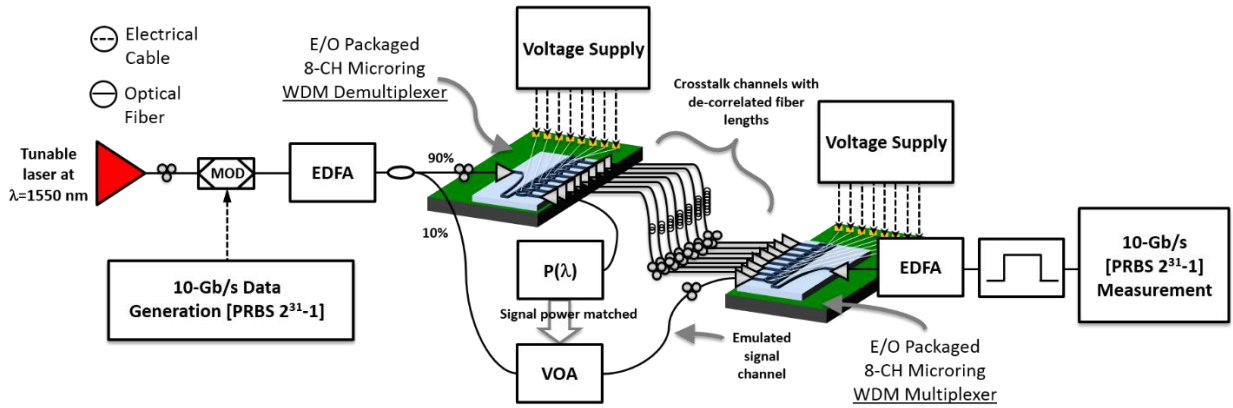

**Figure 4**     Experimental diagram, depicting electrical and optical connections and components used for emulated characterization of the full proposed switching architecture. Voltage supplies used to induce resonance shifts in microrings on the demultiplexer and multiplexer interfaces consisted of digital-to-analog converters controlled by a single FPGA to ensure switching synchronicity. EDFA –erbium doped fiber amplifier; MOD- modulator

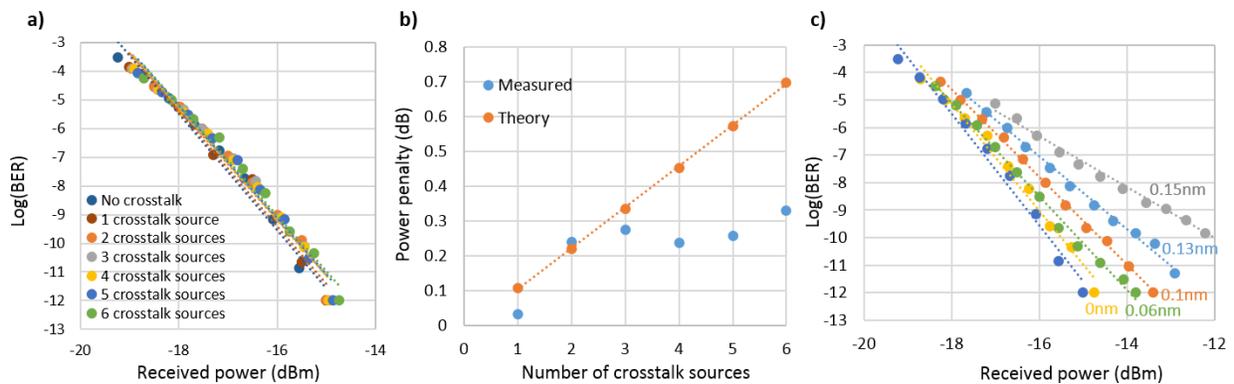

**Figure 5**     **a)** The measured BER of the signal without crosstalk and with progressively increasing from 1 to 6 the number of channels adding crosstalk power to the signal **b)** The measured power penalty, extracted from the BER curves from **a)** and the corresponding theoretical bound **c)** The BER curves for detuned from the resonance microring ; the detuning results in lower signal power but the crosstalk power remains the same.

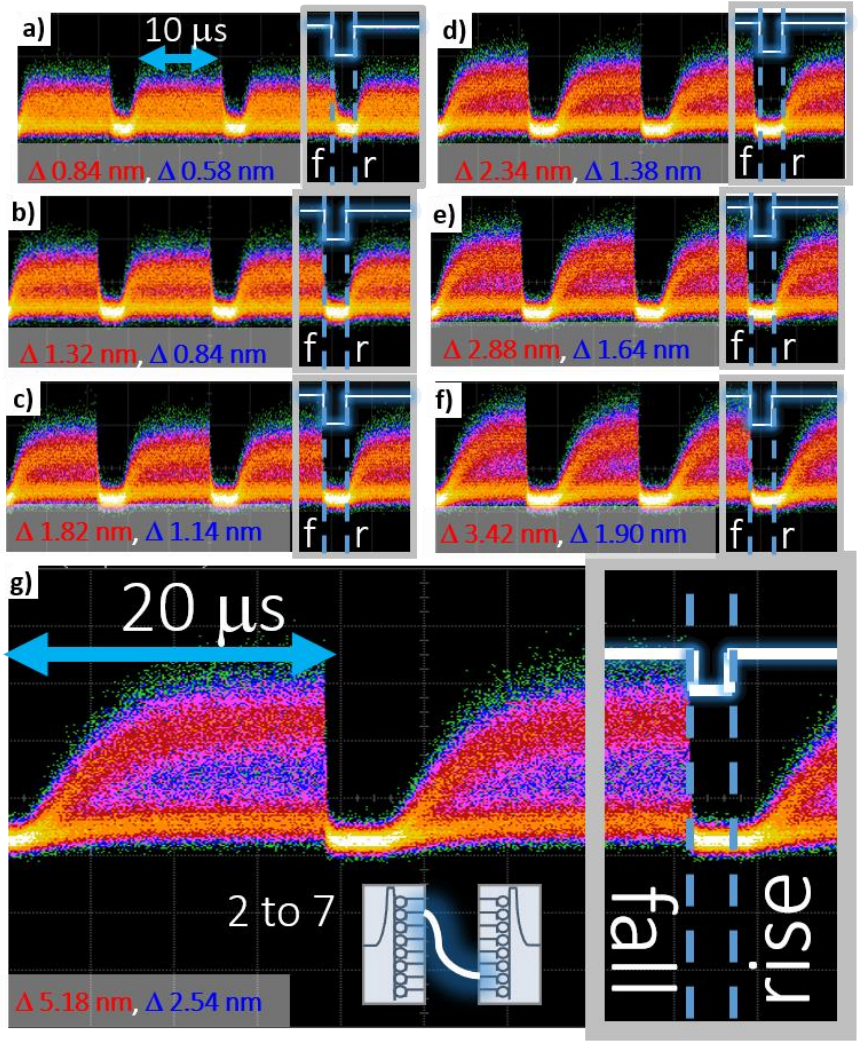

**Figure 6**    Demux outgoing port 2 to mux ingoing port 7 switching characteristics with respect to red and blue resonance shifts of each microring on each PIC, respectively; **(a)** depicts demux microring 2 red shifted and mux microring 7 blue shifted of 0.84 nm and 0.58 nm, respectively; **(b)** microring 2 red shift 1.32 nm and microring 7 blue shift 0.84 nm; **(c)** microring 2 red shift 1.82 nm and microring 7 blue shift 1.14 nm; **(d)** microring 2 red shift 2.34 nm and microring 7 blue shift 1.38 nm; **(e)** microring 2 red shift 1.88 nm and microring 7 blue shift 1.64 nm; **(f)** microring 2 red shift 3.42 nm and microring 7 blue shift 1.90 nm; **(g)** microring 2 red shift 5.18 nm and microring 7 blue shift 2.54 nm. White lines labeled by rise [r] and fall [f] indicate the electrical actuation signal whose associated rise and fall times were on the order of picoseconds, and could not be measured on the same timescale as the optical switching.

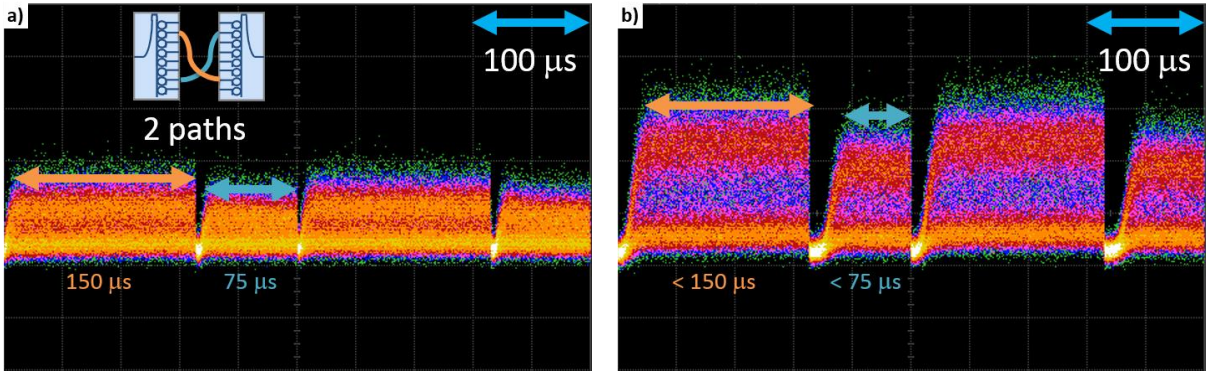

**Figure 7** Multipath FPGA-controlled switching on the microsecond scale showing switching between the paths established by demux microring 2 connected to mux microring 7, and by demux microring 2 connected to mux microring 7; **(a)** shows switching between two chip-to-chip paths when tuning microring resonances at distances corresponding to Figure 6a; **(b)** shows switching between two chip-to-chip paths when tuning microring resonances at distances corresponding to Figure 6g. The path holding times are configured for 300 microseconds on 2 to 7 and 150 microseconds on 7 to 2 in both (a) and (b).

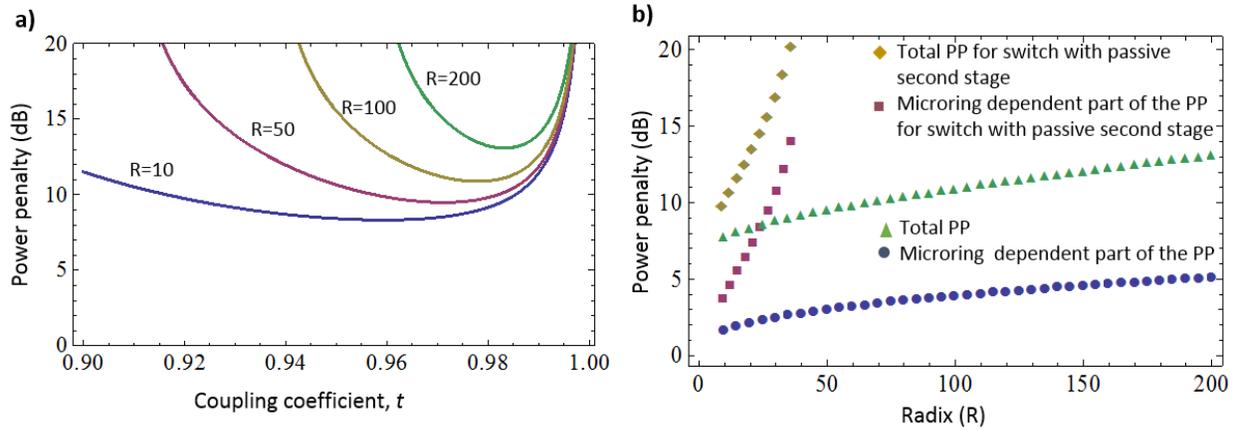

**Figure 8** **(a)** Calculated total power penalty versus the coupling coefficient t for different switch radices (R); **(b)** Calculated power penalties versus the switch radix (number of input (output) ports). The blue dots and red squares are the rings' dependent part of the power penalty for the proposed design and for the design without a second multiplexer chip correspondingly. The green triangles and yellow rhombs show the total power penalty for the corresponding architectures.